\documentclass[aps,prd,onecolumn,preprintnumbers,notitlepage,superscriptaddress,
nofootinbib,amsfont,amssymb,10pt,singlespacing] {revtex4-1}
\usepackage{amsmath,bm}
\usepackage{times}
\usepackage{braket}
\usepackage{color,graphicx}
\usepackage{slashed}
\usepackage{hyperref}
\usepackage{graphics}
\usepackage{subfigure}
\usepackage{multirow,makecell}

\newcommand{\beq}{\begin{eqnarray}}
\newcommand{\eeq}{\end{eqnarray}}

\def\lsim{ {\ \lower-1.2pt\vbox{\hbox{\rlap{$<$}\lower6pt\vbox{\hbox{$\sim$}
}}}\ } }
\def\gsim{ {\ \lower-1.2pt\vbox{\hbox{\rlap{$>$}\lower6pt\vbox{\hbox{$\sim$}
}}}\ } }
\definecolor{Red}{rgb}{1.,0.,0.}

\definecolor{Blue}{rgb}{0.,0.,1.}

\definecolor{nicered}{rgb}{0.7,0.1,0.1}
\definecolor{nicegreen}{rgb}{0.1,0.5,0.1}

\hypersetup{colorlinks,citecolor=nicegreen,linkcolor=nicered}

\begin{document}

\title{
Study of $B_{c}$$\rightarrow$ $\psi(2S)$$K$, $\eta_{c}(2S)$$K$, $\psi(3770)$$K$ decays with perturbative QCD approach}
%%%==================================================================

\author{Feng-Bo~Duan}
\email[Electronic address:]{dfbdfbok@163.com}
\affiliation{School of Physical Science and Technology,
 Southwest University, Chongqing 400715, China}

\author{Xian-Qiao~Yu}
\email[Electronic address:]{yuxq@swu.edu.cn}
\affiliation{School of Physical Science and Technology,
Southwest University, Chongqing 400715, China}

\date{\today}

%%%%%%%%%%%%%%%%%%%%%%%%%%%%%%%%%%%%%%%%%%%%%%%%%%%%%%%%%%%%%%%%%%
\begin{abstract}

We study the $B_{c}$$\rightarrow$$\psi(2S)$K, $\eta_{c}(2S)$K, $\psi(3770)$K decays with perturbative QCD approach (pQCD) based on $k_T$ factorization. The new orbitally excited charmonium distribution amplitudes $\psi(1^{3}D_{1})$ based on the Schr\"{o}dinger wave function of the $n=1$, $l=2$ state for the harmonic-oscillator potential are employed.
By using the corresponding distribution amplitudes, we calculate the branching ratio of $B_{c}$$\rightarrow$$\psi(2S)$K, $\eta_{c}(2S)$K, $\psi(3770)$K decays and
the form factors $A_{0,1,2}$ and $V$ for the transition $B_{c}$$\rightarrow$$\psi(1^{3}D_{1})$. We obtain
the branching ratio of both $B_{c}$$\rightarrow$$\psi(2S)$K and $B_{c}$$\rightarrow$$\eta_{c}(2S)$K are at the order of $10^{-5}$. The effects of two sets of the S-D mixing angle $\theta=-12^{\circ}$ and $\theta=27^{\circ}$
for the decay $B_{c}$$\rightarrow$$\psi(3770)$K are studied firstly in this paper. Our calculations show that the branching ratio of the decay $B_{c}$$\rightarrow$$\psi(3770)$K can be raised from the order of $10^{-6}$ to the order of $10^{-5}$ at the mixing angle $\theta=-12^{\circ}$, which can be tested by the running LHC-b experiments.

\end{abstract}
%%%%%%%%%%%%%%%%%%%%%%%%%%%%%%%%%%%%%%%%%%%%%%%%%%%%%%%%%%%%%%

\pacs{13.25.Hw, 11.10.Hi, 12.38.Bx}
%\preprint{\footnotesize JSNU-PHY-TH-2015}
\maketitle

%
%%%
%%%%%%%%%%%%%%%%% I. INTRODUCTION %%%%%%%%%%%%%%%%%%%%%%%%%%%%%%%%
%%%
%

\section{Introduction}

The detailed study of B meson decay can provides a good chance for testing the Standard Model(SM), searching new physics signals beyond SM\cite{r1}. The meson $B_{c}$,
being the heaviest ground pseudoscalar meson, has been observed for the first time via the decay $B_{c}$$\rightarrow$$J/\psi$$\ell$$\nu$ in 1.8 TeV $P\bar{P}$ collisions through the CDF detector at the Fermilab Tevatron in 1998\cite{r2}. Because it is indifferent to the strong and electromagnetic interactions, it can decay only through the weak interaction.  The meson $B_{c}$ has rich decay channels\cite{Xiao}, for either of its component, i.e., b or c quarks, can decay individually. Due to this innate advantage, it can also provide a very ideal platform to study weak decays of heavy quarks.

Recently, the decay $B_{c}$$\rightarrow$$\psi(2S)$$\pi$ had been updated by the LHC-b Collaboration accompanied by the measured ratio of the branching fractions \cite{r3} since
its first observation in 2013
\begin{equation}
\begin{split}
\frac{BR({B_{c}\rightarrow\psi(2S)\pi})}{BR({B_{c}\rightarrow J/\psi\pi})}&=0.268\pm0.032({\rm stat})\\
&\pm0.007({\rm syst})\pm0.006({\rm BF})
\end{split}
\end{equation}
The first uncertainty is statistical, the second is systematic, and the last term indicates the uncertainty from Br($\psi(2S)$$\rightarrow$$\mu^+\mu^-$)/Br \\ ($J/\psi$$\rightarrow$$\mu^+\mu^-$). Here $\psi(2S)$ is the first radially excited charmonium meson. About the vector charmonium meson $\psi(2S)$ contained in a $B_{c}$ decays, it has been studied in various approaches. For example, in Ref.\cite{r4}, the authors calculated the branching ratios for the $B_{c}$$\rightarrow$$\psi(2S)$$X$ by means of the modified harmonic-oscillator wave function based on the light front quark model; in Ref.\cite{r5}, the authors used ISGW2 quark model to research the production of radially excited charmonium mesons in $B_{c}$ decays; the relativistic (constituent) quark model, the potential model, the QCD relativistic potential model, and the improved instantaneous BS equation and Mandelstam approach were adopted in Refs.\cite{r6,r7,r8,r9,r10}, respectively. However, it is regret to tell that all of these computations are based on a so-called naive factorization assumption, with various form factor inputs. There are also some uncontrolled large theoretical errors with quite different numerical results. Constraining by the unreliability of their models, most of them cannot give any theoretical error estimates. In the work\cite{r11}, the authors successfully used pQCD\cite{r12} approach to study the S-wave ground state charmonium decays of $B_{c}$ meson based on the harmonic-oscillator wave functions for the charmonium 1S states. In this paper, we also take harmonic-oscillator wave functions as the approximate wave function of both 2S and 1D charmonium states to study the $B_{c}$$\rightarrow$$\psi(2S)$K, $\eta_{c}(2S)$K, $\psi(3770)$K decays.
 Here 1D charmonium state is the component of $\psi(3770)$ resonance. $\psi(3770)$, the lowest-lying charmonium state above the open-charm $D\bar{D}$ threshold, is of great interest in quarkonium physics. The rate of decay mode $B$$\rightarrow$$\psi(3770)$K, observed in the Belle Collaboration\cite{r13}, is surprisingly large. It might seemingly indicate that the $\psi(3770)$ is mainly the vector charmonium state $\psi(1^{3}D_{1})$ with a small admixture of vector charmonium state $\psi(2^{3}S_{1})$. It is expected to be expressed as
\begin{equation}
\psi(3770)=\cos\theta|c\bar{c}(1^{3}D_{1})\rangle-\sin\theta|c\bar{c}(2^{3}S_{1})\rangle.
\end{equation}
Here, the S-D mixing angle $\theta$ arises from the ratio of the leptonic decay widths of $\psi(3686)$ and $\psi(3770)$ \cite{r14}. Calculations from nonrelativistic potential model
provide two sets of mixing scheme: $\theta=-(12\pm2)^{\circ}$ or $\theta=(27\pm2)^{\circ}$ \cite{Kuang,Rosner,r17}. In Ref\cite{r18}, the authors have used the
light-cone QCD sum rules to calculate the form factors $A^{B_{c}\rightarrow\psi(1^{3}D_{1})}_{0,1,2}$ and $V^{B_{c}\rightarrow\psi(1^{3}D_{1})}$. In view of the simple analysis, we
will calculate the form factors $A^{B_{c}\rightarrow\psi(1^{3}D_{1})}_{0,1,2}$ and $V^{B_{c}\rightarrow\psi(1^{3}D_{1})}$ by using the pQCD approach in this work.

This paper is organized as follows. In Sect.II, we describe the theoretical framework and the wave function for the radially excited charmonium mesons $\psi(2S)$, $\eta_{c}(2S)$ and
the orbitally excited charmonium state $\psi(1^{3}D_{1})$. In Sect.III, we present the corresponding form factor expressions for the transition $B_{c}\rightarrow\psi(1^{3}D_{1})$. The
decay amplitudes for the two-body decays $B_{c}\rightarrow X_{c\bar{c}}K(X=\psi(2S), \eta_{c}(2S), \psi(3770))$ are computed by employing the pQCD approach in Sect.IV. The numerical
results and several points of discussions are presented in Sect.V. Finally we will finish this paper with a brief summary.

%
%%%
%%%%%%%%%%%%%%%%% II. Theoretical frame and the wave function %%%%%%%%%%%%%%%%%%%%%%%%%%%%%%%%
%%%
%
\section{Theoretical frame and the wave function}\label{sec:pert}

The $B_{c}$($X_{c\bar{c}}$) meson momentum $p_{1}(p_{2})$ and the light quarks momentum $\emph{k}_{i}$ included in each meson are writen in the
light-cone coordinates as

$p_{1}=\frac{M_{B_{c}}}{\sqrt 2}(1,1,0_{\top})$,                $k_{1}=x_{1}p_{1}+(0,0,k_{1\top})$

$p_{2}=\frac{M_{B_{c}}}{\sqrt 2}(1,r^2_{2},0_{\top})$,           $k_{2}=x_{2}p_{2}+(0,0,k_{2\top})$

$p_{3}=\frac{M_{B_{c}}}{\sqrt 2}(0,1-r^2_{2},0_{\top})$,           $k_{3}=x_{3}p_{3}+(0,0,k_{3\top})$ \\
with $r_{2}=\frac{M_{X_{c\bar{c}}}}{M_{B_{c}}}$. The $K$ meson momentum $p_{3}$=$p_{1}$-$p_{2}$. The polarization vectors of the vector mesons $\psi(2S)$ and $\psi(3770)$ are given as\\
$\epsilon_{L}=\frac{1}{\sqrt2}(\frac{1}{r_{2}},-r_{2},0)$, $\epsilon_{\top}=(0,0,1_{\top})$

For the decay $B_{c}$$\rightarrow$$X_{c\bar{c}}$$K$, the relevant effective Hamiltonian is written as\cite{r19}
\begin{equation}
{\cal H}_{eff}=\frac{G_{F}}{\sqrt2}V_{us}V^*_{cb}\{C_{1}(\mu)O_{1}(\mu)+C_{2}(\mu)O_{2}(\mu)\}+H.c.,
\end{equation}
with the Cabibbo-Kobayashi-Maskawa (CKM) matrix elements $V_{us}$ and $V^*_{cb}$, and the local four-fermion operators
\begin{equation}
O_{1}=(\bar{b}_{i}c_{j})_{V-A}(\bar{u}_{j}s_{i})_{V-A},O_{2}=(\bar{b}_{i}c_{i})_{V-A}(\bar{u}_{j}s_{j})_{V-A}.
\end{equation}
here i and j denote SU(3) color indices. C($\mu$) is the Wilson coefficient estimated at renormalization scale $\mu$. Apparently, because there are not any two of the same quarks in the four-quark operators, penguin diagrams can not contribute. Therefore there will be no $CP$ violation in the decays of $B_{c}$$\rightarrow$$X_{c\bar{c}}$$K$ within the standard model.

In the pQCD theoretical frame, the decay amplitude can be decomposed as the convolution\cite{r20,r21,r22,r23}
\beq
\begin{split}
{\cal A}(B_{c}\rightarrow &M_{2}M_{3})\sim \int d^4\emph{k}_{1}d^4\emph{k}_{2}d^4\emph{k}_{3}\\
&Tr[\phi_{B_{c}}(\emph{k}_{1})\phi_{X}(\emph{k}_{2})\phi_{K}(\emph{k}_{3})
H(\emph{k}_{1},\emph{k}_{2},\emph{k}_{3},t)]e^{-S(t)},\label{eq:exp5}
\end{split}
\eeq

where $Tr$ refers to the trace over Dirac and color indices; the function $H(\emph{k}_{1}, \emph{k}_{2}, \emph{k}_{3},t)$ describes the so-called hard scattering kernels, which is scale dependent but can be perturbative calculated fortunately. The function $\phi_{B_{c}}(\emph{k}_{1})$, $\phi_{X}(\emph{k}_{2})$ and $\phi_{K}(\emph{k}_{3})$ denote hadron wave functions, which play the role of absorbing the infrared divergence. The Sudakov factors $S(t)$ arise from both $\emph{k}_{\top}$ and threshold resummation, aiming to avoid the end-point singularity.

In our calculations, the distribution amplitude of realistic model for hadron $B_{c}$ can be found in Ref\cite{r241,r242}
\begin{equation}
\phi_{B_{c}}(\emph{u})=6\emph{u}(1-\emph{u})[1+\sum^{\infty}_{\emph{n}=1}\emph{a}_{\emph{n}}(\mu)\emph{C}^{3/2}_{\emph{n}}(2\emph{u}-1)],
\end{equation}
where $\emph{C}^{3/2}_{1}(x)=3x$, $\emph{C}^{3/2}_{2}(x)=\frac{3}{2}(5x^{2}-1)$, and $\emph{a}_{\emph{n}}$ denotes the Gegenbauer moments. In Ref\cite{r241}, the authors have calculated the relativistic corrections of Gegenbauer moments and found that they are comparable with the next to leading order radiative corrections, and they have also given the total correction values for the first two Gegenbauer moments $\emph{a}_{1}$ and $\emph{a}_{2}$, which contain leading order contribution, one-loop QCD radiative corrections and relativistic corrections.

For the light pseudoscalar meson kaon, the wave function can generally be defined as\cite{r26}

\begin{equation}
\phi(p,x,\xi)=\frac{i}{2N_{c}}\gamma_{5}[/\kern-0.6em p\phi^A_{\emph{k}}(x)+m_{0}\phi^P_{\emph{k}}(x)+\xi m_{0}(/\kern-0.6em n /\kern-0.6em \upsilon -1)\phi^T_{\emph{k}}(x)]
\end{equation}
We adopt the distribution amplitudes $\phi^{A,\emph{P},T}_{\emph{k}}$ from Ref.\cite{r27,r28}:

\begin{equation}
\phi^A_{\emph{k}}(\emph{x})=\frac{6f_{\emph{K}}}{2\sqrt{2N_{c}}}\emph{x}(1-\emph{x})[1+0.15t+0.405(5t^2-1)]
\end{equation}

\begin{equation}
\phi^\emph{P}_{\emph{k}}(\emph{x})=\frac{f_{\emph{K}}}{2\sqrt{2N_{c}}}[1+0.106(3t^2-1)-0.148(3-30t^2+35t^4)/8]
\end{equation}

\begin{equation}
\phi^T_{\emph{k}}(\emph{x})=\frac{f_{\emph{K}}}{2\sqrt{2N_{c}}}t[1+0.1581(5t^2-3)]
\end{equation}
with $t=1-2\emph{x}$. Here the wave function $\phi^{A}_{\emph{K}}$ refers to the twist-2 distribution amplitude, and both $\phi^{\emph{P}}_{\emph{K}}$ and $\phi^{T}_{\emph{K}}$ refer to the twist-3 distribution amplitudes.

In Refs.\cite{r29,r30}, the harmonic-oscillator wave functions have been applied to describe the charmonium ground state $J/\psi$, and the theoretical results agree well with the published experimental data. In Ref\cite{r24}, the authors also adopted the harmonic-oscillator wave functions for the mesons $\psi(2S)$ and $\eta_{c}(2S)$, the ratio $B_{r}$$(B_{c}$$\rightarrow$$\psi(2S)\pi)$/$B_{r}$$(B_{c}$$\rightarrow$$J/\psi\pi)$=$0.29^{+0.17}_{-0.11}$ they got are very close to the experimental data $0.268\pm0.032\pm0.007$\cite{r3}. It pushes us to try to predict the branching ratio for the $B_{c}$$\rightarrow$$\psi(3770)$$K$
decays, for which no one has yet made a theoretical prediction. As what mentioned before, the meson $\psi(3770)$ is 2S-1D mixing state. The pure 2S state has been mentioned above. The pure 1D state indicates the principal quantum number $n$=1 and the orbital angular momentum $l$=2, which means that it is only a angular excitation state. We are going to
describe it by using harmonic-oscillator wave functions for $n$=1, $l$=2 Schr\"{o}dinger state.

For the wave function of the vector charmonium states $\psi(2S)$, $\psi_{c\bar{c}}(1D)$, we refer to the vector mesons $\omega$, $\rho$ and $D^{*}$ in Refs.\cite{r32,r33}, and the wave function of the pseudoscalar meson $\eta_{c}(2S)$ gets the same access to $\eta_{c}(1S)$\cite{r29}.
\begin{equation}
\begin{split}
\langle\psi &(p_{2},\epsilon_{L})|\bar{c}_{\alpha}(0)c_{\beta}(z)|0\rangle \\
&=\frac{1}{\sqrt{2N_{c}}}\int d^{4}\emph{k}_{2}e^{+i\emph{k}_{2}\cdot\emph{z}}/\kern-0.6em \epsilon_{L}[\emph{m}_{\psi_{c\bar{c}}(1D)}\phi^{\emph{L}}_{\psi}(\emph{k}_{2})
+/\kern-0.6em \emph{p}_{2}\phi^{\emph{t}}_{\psi}(\emph{k}_{2})]_{\beta\alpha}
\end{split}
\end{equation}

\begin{equation}
\begin{split}
\langle\psi &(p_{2},\epsilon_{T})|\bar{c}_{\alpha}(0)c_{\beta}(z)|0\rangle\\
&=\frac{1}{\sqrt{2N_{c}}}\int d^{4}\emph{k}_{2}e^{+i\emph{k}_{2}\cdot\emph{z}}/\kern-0.6em \epsilon_{T}\phi^{\emph{T}}_{\psi}(\emph{k}_{2})[\emph{m}_{\psi_{c\bar{c}}(1D)}
+/\kern-0.6em \emph{p}_{2}]_{\beta\alpha}
\end{split}
\end{equation}

\begin{equation}
\begin{split}
\langle\eta_{c}&(2S)(p_{2})|\bar{c}_{\alpha}(0)c_{\beta}(z)|0\rangle \\
&=\frac{-i}{\sqrt{2N_{c}}}\int d^{4}\emph{k}_{2}e^{+i\emph{k}_{2}\cdot\emph{z}}{[\gamma_{5}/\kern-0.6em \emph{p}_{2}\phi^{\upsilon}(\emph{k}_{2})
+\gamma_{5}\emph{m}\phi^{\emph{s}}(\emph{k}_{2})]}_{\beta\alpha}
\end{split}
\end{equation}
where $p_{2}$ plays the part of the momentum of the charmonium mesons $\psi(2S)$, $\psi(1^3D_{1})$, or $\eta_{c}(2S)$ and $m$ is their corresponding mass. The $\epsilon^{\emph{L}(\emph{T})}$ means the longitudinal (transverse) polarization vector. Here the functions $\phi^{\emph{L}}_{\psi}$, $\phi^{\emph{T}}_{\psi}$ and $\phi^{\upsilon}$ pertain to twist-2 distribution amplitudes, and both the $\phi^{\emph{t}}_{\psi}$ and $\phi^{\emph{s}}$ pertain to the twist-3 distribution amplitudes.
Their distribution amplitudes of the asymptotic models for the radially excited charmonium mesons $\psi(2S)$ and $\eta_{c}(2S)$ have been studied in Ref.\cite{r24}. We are going to focus on the distribution amplitude of the asymptotic model for $\psi(1^3D_{1})$ state as follows.

First of all, we give the isochronous Schr\"{o}dinger equation based on the harmonic-oscillator potential as
\begin{equation}
\psi_{1D}(\text{r})\propto(\alpha{r})^2e^{-\frac{{\alpha}^2 r^2}{2}}Y_2m(\theta,\varphi)£¬
\end{equation}
where $Y_2m(\theta,\varphi)$ is the spherical harmonic function. ${\alpha}^2=\frac{m_{c}\omega}{2}$ and $\omega$ is the harmonic vibration frequency.

In order to get its function in the momentum space, we apply the Fourier transform to it,
\begin{equation}
\psi_{1D}(\overrightarrow{\emph{k}})=\int d^{3}\overrightarrow{\emph{r}}e^{-i\overrightarrow{\emph{r}}\cdot\overrightarrow{\emph{k}}}\psi_{1D}(\text{r}) \propto(\overrightarrow{\emph{k}}^2-3\emph{k}^2_\emph{z})e^{-\frac{\overrightarrow{\emph{k}}^2}{2{\alpha}^2}},
\label{eq:exp15}
\end{equation}
where $\emph{k}$ represents three-dimensional momentum.

Taking the substitution ansatz\cite{r34,r35}
\begin{equation}
\overrightarrow{\emph{k}_\bot}\rightarrow \overrightarrow{\emph{k}_\bot},\emph{k}_\emph{z}\rightarrow (\emph{x}-\overline{\emph{x}})\frac{\emph{m}_0}{2},\emph{m}^2_0=\frac{\emph{m}^2_c +\overrightarrow{\emph{k}_\bot}^2}{\emph{x}\overline{\emph{x}}},
\end{equation}
where $\overline{\emph{x}}=1-\emph{x}$, and $\emph{x}$ is the momentum fraction associated with one of the partons. Now we will write down the relationship formula
\begin{equation}
\emph{k}^2 \rightarrow \frac{\emph{k}^2_\bot +(\emph{x}-\overline{\emph{x}})^2\emph{m}^2_c}{4 \emph{x}\overline{\emph{x}}},
\end{equation}
then the wave function (\ref{eq:exp15}) is replaced by
\begin{equation}
\begin{split}
&\psi_{1D}(\overrightarrow{\emph{k}})\rightarrow \psi_{1D}(\emph{x},\overrightarrow{\emph{k}_\bot})\\
&\propto(\frac{\emph{k}^2_\bot +(\emph{x}-\overline{\emph{x}})^2\emph{m}^2_c}{4 \emph{x}\overline{\emph{x}}} -3 \frac{(\emph{k}^2_\bot +\emph{m}^2_c)(\emph{x}-\overline{\emph{x}})^2}{4 \emph{x}\overline{\emph{x}}})e^{-\frac{\emph{k}^2_\bot +(\emph{x}-\overline{\emph{x}})^2\emph{m}^2_c}{8 \emph{x}\overline{\emph{x}}\alpha^2}}
\end{split}
\end{equation}

Now we're going to convert the transverse momentum $\emph{k}_\bot$ to its conjugate variable $\emph{b}$, the oscillator wave function $\psi_{1D}(\emph{x},\emph{b})$ can be written as

\begin{equation}
\begin{split}
\psi_{1D}(\emph{x},\text{\emph{b}})\sim &\int  \text{d}^2 \text{\emph{k}}_\bot e^{-i\text{\emph{b}}\cdot\text{\emph{k}}_\bot} \psi_{1D}(\emph{x},\text{\emph{k}}_\bot)\\
&\propto\emph{x}\overline{\emph{x}}{\cal I}(\emph{x})e^{-\emph{x}\overline{\emph{x}}\frac{\emph{m}_c}{\omega}
[\omega^2\emph{b}^2+(\frac{\emph{x}-\overline{\emph{x}}}{2\emph{x}\overline{\emph{x}}})^2]},
\end{split}
\end{equation}
with
\begin{equation}
{\cal I}(\emph{x})=(\frac{1}{\emph{x}\overline{\emph{x}}}-\emph{m}\omega \emph{b}^2)(6\emph{x}^4-12\emph{x}^3+7\emph{x}^2-\emph{x})-\frac{\emph{m}_c(1-2\emph{x})^2}{4\omega \emph{x}\overline{\emph{x}}}.
\end{equation}
The modified wave functions can be given as \begin{equation}
\psi_{1D}(\emph{x},\text{\emph{b}})\propto \Phi^{asy}(\emph{x}){\cal I}(\emph{x})e^{-\emph{x}\overline{\emph{x}}\frac{\emph{m}_c}{\omega}
[\omega^2\emph{b}^2+(\frac{\emph{x}-\overline{\emph{x}}}{2\emph{x}\overline{\emph{x}}})^2]}
\end{equation}
with the $\Phi^{asy}(\emph{x})$ being the asymptotic models\cite{r35}. We then obtain the distribution amplitudes for the orbitally excited charmonium state $\psi(1^3D_{1})$
\begin{equation}
\psi^{\emph{L},\emph{T}}_{1D}(\emph{x},\text{\emph{b}})= \frac {\emph{f}_{1D}}{2\sqrt{2N_{c}}}N^{\emph{L},\emph{T}} \emph{x}\overline{\emph{x}} {\cal I}(\emph{x})e^{-\emph{x}\overline{\emph{x}}\frac{\emph{m}_c}{\omega}
[\omega^2\emph{b}^2+(\frac{\emph{x}-\overline{\emph{x}}}{2\emph{x}\overline{\emph{x}}})^2]}
\label{eq:exp22}
\end{equation}

\begin{equation}
\psi^{\emph{t}}_{1D}(\emph{x},\text{\emph{b}})= \frac {\emph{f}_{1D}}{2\sqrt{2N_{c}}}N^{\emph{t}} ({\emph{x}-\overline{\emph{x}}})^2{\cal I}(\emph{x})e^{-\emph{x}\overline{\emph{x}}\frac{\emph{m}_c}{\omega}
[\omega^2\emph{b}^2+(\frac{\emph{x}-\overline{\emph{x}}}{2\emph{x}\overline{\emph{x}}})^2]},
\label{eq:exp23}
\end{equation}
with the normalization conditions:

\begin{equation}
\int^1_0 \psi^{\emph{i}}_{1D}(\emph{x},{0})dx= \frac {\emph{f}_{1D}}{2\sqrt{2N_{c}}}
\end{equation}
where $N_c=3$ is the color number, $N^i$(i=$\emph{L}, \emph{T}, \emph{t}$) are the normalization constants, and $\emph{f}_{1D}$=47.8MeV\cite{r18} is the decay constant of the orbitally excited $\psi({1^3D_1})$ state. Both the wave functions Eq.~(\ref{eq:exp22}) and Eq.~(\ref{eq:exp23}) are symmetric under $\emph{x}\leftrightarrow \overline{\emph{x}}$.

To calculate the decay branching ratio of the model $B_{c}\rightarrow \psi(1^3D_1)+\emph{K}$, it is the frequency of oscillations $\omega_{1D}$ that can not be determined easily.
For light $q\bar{q}$ systems, the best value of oscillation frequency from spectroscopy and decays is 0.379 GeV, but for heavy $Q\bar{Q}$ states, it is a bit larger and also has a range $\omega$=0.4$\sim$0.6 GeV in the literature(see Ref.\cite{Anwar} and references therein). According to the quark model theory, studies show that the effective oscillatory parameter $\omega$ for higher $\emph{c}\overline{\emph{c}}$ multiplets is smaller than the corresponding lower ones\cite{Anwar}, that is based on the fact that the excited states have large spatial extensions. For example, in Ref.\cite{r29}, the authors tried to take the ground charmonium state $J/\psi$ frequency from 0.5GeV to 0.8 GeV, in Ref.\cite{r24}, the authors tried the value $\omega=0.2$ GeV for the first radially excited charmonium mesons $\psi(2S)$ and $\eta_{c}(2S)$. Although it's still difficult to define precisely what the $\omega$ for $\psi(1^3D_1)$ state is, we will try to adopt $\omega$=0.35$\sim$0.55GeV for $\psi(1^3D_1)$ state in this work, that is reasonable based on the above theoretical analysis.

%%%--=================================================================
%%%=====      Form factors         ==========================
%%5===================================================================
\section{$B_{c}\rightarrow \psi({1^3D_{1}})$ Form factors} \label{sec:numer}

Based on the pQCD theoretical framework, the form factors of the transition $B_{c}\rightarrow \psi({1^3D_{1}})$ is similar to that of $B_{c}\rightarrow J/\psi$ which can be defined as \cite{r37,r38}

\begin{equation}
\begin{split}
\langle\psi(1^3D_1)&(p_{2},\epsilon_{T})|\bar{c}\gamma_{\mu}\emph{b}|B_{c}(\emph{p}_1)\rangle\\
&=\frac{2iV(\emph{q}^2)}{\emph{M}_{B_{c}}+\emph{m}_{\psi(1D)}}\epsilon^{\mu\nu\rho\sigma}\epsilon^*_{\nu}\emph{p}_{2\rho}\emph{p}_{2\sigma}\\
\end{split}
\end{equation}

\begin{equation}
\begin{split}
\langle\psi &(1^3D_1)(p_{2},\epsilon_{T})|\bar{c}\gamma_{\mu}\gamma_{5}\emph{b}|B_{c}(\emph{p}_1)\rangle\\
&=\frac{\epsilon^*\cdot \emph{q}}{\emph{q}^2}2\emph{m}_{\psi(1D)}\emph{q}^{\mu}A_0^{B_{c}\rightarrow \psi(1^3D_{1})}(\emph{q}^2)+(\emph{M}_{B_{c}}+\emph{m}_{\psi(1D)})\\ & A_1^{B_{c}\rightarrow \psi(1^3D_{1})}(\emph{q}^2)[\epsilon^{*\mu}-\frac{\epsilon^*\cdot \emph{q}}{\emph{q}^2}\emph{q}^{\mu}]-\frac{\epsilon^*\cdot\emph{q}}{\emph{M}_{B_{c}}+\emph{m}_{\psi(1D)}}\\
&[(\emph{p}_1+\emph{p}_2)^{\mu}-\frac{\emph{M}^2_{B_c}-\emph{m}^2_{\psi(1D)}}{\emph{q}^2}\emph{q}^{\mu}]A_2^{B_{c}\rightarrow \psi(1^3D_{1})}(\emph{q}^2)\\
\end{split}
\end{equation}
where $\emph{q}$=$\emph{p}_1-\emph{p}_2$ is the momentum transfer and $\epsilon^*$ represents the polarization vector of the $\psi(1^3D_1)$ charmonium state. $V^{B_{c}\rightarrow \psi(1^3D_1)}$ and $A^{B_{c}\rightarrow \psi(1^3D_1)}_{0,1,2}$ are the transition form factors. Furthermore, in the large-recoil limit, i.e. $\emph{q}^2=0$, we have
\begin{equation}
A_0(0)=\frac{1+\emph{r}_{2}}{2\emph{r}_{2}}A_1(0)-\frac{1-\emph{r}_{2}}{2\emph{r}_{2}}A_2(0),
\end{equation}
where $\emph{r}_{2}=\frac{m_{\psi(1D)}}{M_{B_c}}$.

\begin{figure}[htbp]
\centering
\begin{tabular}{l}
\includegraphics[width=0.75\textwidth]{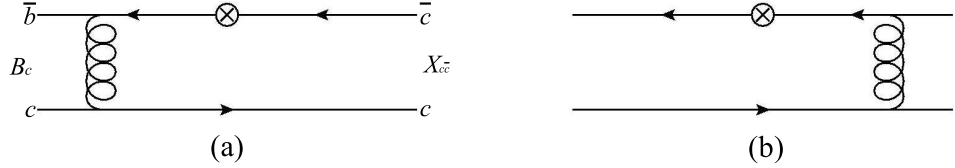}
\end{tabular}
\caption{Feynman diagrams contributing to the $B_{c}\rightarrow \emph{X}_{c\bar c}$ form factors}
  \label{fig:fig1}
\end{figure}

Based on the single gluon exchange, the lowest-order diagrams (a) and (b) in Fig.~\ref{fig:fig1}. contribute to the transition form factor for the $B_{c}\rightarrow \psi({1^3D_{1}})$ at the maximally recoiling point ($q^2=0$). Our predictions of the form factors are collected in Table I and are compared with the results from the light cone QCD sum rule \cite{r18}.

In the pQCD theory scheme, the expressions for the form factor $A^{B_{c}\rightarrow\psi(1^{3}D_{1})}_{0,1,2}$ and $V^{B_{c}\rightarrow\psi(1^{3}D_{1})}$ are written directly as
\begin{equation}
\begin{split}
A_{0}&=-8\pi M^{2}_{B_{c}}\emph{C}_{\emph{F}}\int^{1}_{0}\text{d}\emph{x}_{1}\text{d}\emph{x}_{2}\int^\infty_{0}\emph{b}_{1}\emph{b}_{2}\text{d}\emph{b}_{1}\text{d}\emph{b}_{2}\phi_{B_{c}}(\emph{x}_{1})\\
& \times \{  [(1-x_{2}-2r_{\emph{b}})\phi^{\emph{L}}(\emph{x}_{2},b_{2})+(2\emph{x}_{2}-2+r_{\emph{b}})r_{2}\phi^{t}(\emph{x}_{2},b_{2})]\\
& \times \emph{E}_{a}(t_{a})H_{a}(\alpha,\beta_{a},b_{1},b_{2})S_{t}(x_{2})+[r^2_{2}(x_{1}-1)-x_{1}]\\
&\times \phi^{\emph{L}}(\emph{x}_{2},b_{2})\emph{E}_{b}(t_{b})H_{b}(\alpha,\beta_{b},b_{1},b_{2})S_{t}(x_{1}) \},
\end{split}
\end{equation}

\begin{equation}
\begin{split}
A_{1}&=-\frac{r_{2}}{1+r_{2}}8\pi M^{2}_{B_{c}}\emph{C}_{\emph{F}}\int^{1}_{0}\text{d}\emph{x}_{1}\text{d}\emph{x}_{2}\int^\infty_{0}\emph{b}_{1}\emph{b}_{2}\text{d}\emph{b}_{1}\text{d}\emph{b}_{2}\phi_{B_{c}}(\emph{x}_{1})\\
& \times \{  [(2-x_{2}-4r_{b}-x_{2}r^2_{2})\phi^{\emph{L}}(\emph{x}_{2},b_{2})\\
& +(r_{b}r_{2}-2r_{2}+4x_{2}r_{2}+\frac{r_{b}}{r_{2}}-\frac{2}{r_{2}})\phi^{t}(\emph{x}_{2},b_{2})]\\
& \times \emph{E}_{a}(t_{a})H_{a}(\alpha,\beta_{a},b_{1},b_{2})S_{t}(x_{2})\\
& -(1+2r_{c}-2x_{1}+r^2_{2})\phi^{\emph{L}}(\emph{x}_{2},b_{2})\\
&  \times \emph{E}_{b}(t_{b})H_{b}(\alpha,\beta_{b},b_{1},b_{2})S_{t}(x_{1}) \},
\end{split}
\end{equation}

\begin{equation}
\begin{split}
A_{2}&=\frac{r_{2}}{1-r_{2}}8\pi M^{2}_{B_{c}}\emph{C}_{\emph{F}}\int^{1}_{0}\text{d}\emph{x}_{1}\text{d}\emph{x}_{2}\int^\infty_{0}\emph{b}_{1}\emph{b}_{2}\text{d}\emph{b}_{1}\text{d}\emph{b}_{2}\phi_{B_{c}}(\emph{x}_{1})\\
& \times \{  [(x_{2}r^2_{2}-x_{2})\phi^{\emph{L}}(\emph{x}_{2},b_{2})
 +(r_{b}r_{2}-2r_{2}-\frac{r_{b}}{r_{2}}+\frac{2}{r_{2}})\\
& \phi^{t}(\emph{x}_{2},b_{2})] \times \emph{E}_{a}(t_{a})H_{a}(\alpha,\beta_{a},b_{1},b_{2})S_{t}(x_{2}) \\
& +(1-2x_{1})(1-r^2_{2})\phi^{\emph{L}}(\emph{x}_{2},b_{2})\\
&  \times \emph{E}_{b}(t_{b})H_{b}(\alpha,\beta_{b},b_{1},b_{2})S_{t}(x_{1}) \},
\end{split}
\end{equation}

\begin{equation}
\begin{split}
V&=-(1+r_{2})8\pi M^{2}_{B_{c}}\emph{C}_{\emph{F}}\int^{1}_{0}\text{d}\emph{x}_{1}\text{d}\emph{x}_{2}\int^\infty_{0}\emph{b}_{1}\emph{b}_{2}\text{d}\emph{b}_{1}\text{d}\emph{b}_{2}\phi_{B_{c}}(x_{1})\\ &\times \phi^{\emph{L}}(\emph{x}_{2},b_{2}) \{  (x_{2}r_{2}+r_{b}-2)\times \emph{E}_{a}(t_{a})H_{a}(\alpha,\beta_{a},b_{1},b_{2})S_{t}(x_{2}) \\
& -r_{2}\times \emph{E}_{b}(t_{b})H_{b}(\alpha,\beta_{b},b_{1},b_{2})S_{t}(x_{1}) \},
\end{split}
\end{equation}
where $r_{b,c}=\frac{m_{b,c}}{M_{B_{c}}}$ and $C_{F}=4/3$ is the group factor of the $SU(3)_{c}$ gauge group. The function $E_{i}(t_{i})$, the hard-scattering kernel function $H_{i}$ and the parametrization factor $S_{t}(x)$ are displayed together in the Appendix.

%%%%%%%%%%%%%%%%%%%%%%%%%%%%%%%%%%%%%%%%%%%%%%%%%%%%%%%%%%%%%%%%%%%%%%%%%

%%%--=================================================================
%%%=====           decay amplitudes    ==========================
%%5===================================================================

\section{The decay amplitudes}
\begin{figure}[htbp]
 \centering
 \begin{tabular}{l}
 \includegraphics[width=0.75\textwidth]{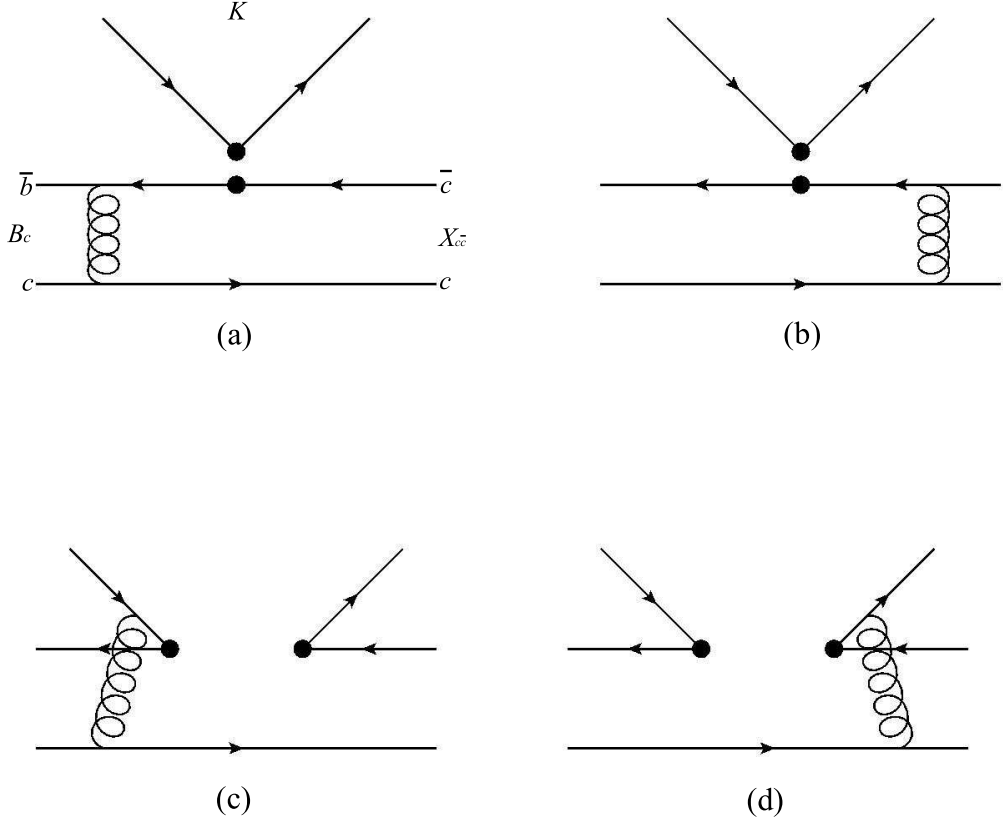}
 \end{tabular}
 \caption{The lowest order Feynman diagrams for the $B_{c}\rightarrow \emph{X}_{c\bar c}$ decays}
   \label{fig:fig2}
 \end{figure}

The decays $B_{c}$$\rightarrow$ $X_{c\bar c}$$K$ are dominated by tree diagrams based on the operator product expansion. Through the analysis in Section 2, there is no pollution from penguins and annihilation diagrams. In the perturbative QCD approach, the Feynman diagrams
are displayed in Fig.~\ref{fig:fig2}, where (a) and (b) are of factorizable topology; (c) and (d) are of nonfactorizable topology. Furthermore, we can directly use Eq.(\ref{eq:exp5}) to give the decay amplitudes
\begin{equation}
{\cal A}(B_{c} \rightarrow \emph{X}_{c\bar c}K)=\frac{G_{F}}{\sqrt{2}}V_{us}V^*_{cb}\sum_{i=a,b,c,d}{\cal A}_{i}
\end{equation}

The explicit expressions for the relative amplitudes ${\cal A}_{i}$ are displayed as follows:

The amplitudes for $B_{c}$$\rightarrow$ $\psi(2S)$$K$ decay,
\begin{equation}
\begin{split}
{\cal A}_{a}=&-8\pi\emph{C}_{\emph{F}}f_{\emph{k}}M^{4}_{B_{c}}(1-r^2_{2})\\
& \times \int^{1}_{0}\text{d}\emph{x}_{1}\text{d}\emph{x}_{2}\int^\infty_{0}\emph{b}_{1}\emph{b}_{2}\text{d}\emph{b}_{1}\text{d}\emph{b}_{2}\phi_{B_{c}}(\emph{x}_{1})\\
& \times \emph{E}_{a}(t_{a})\emph{C}_{a}(t_{a})H_{a}(\alpha,\beta_{a},b_{1},b_{2})S_{t}(x_{2})\\
& \times [(1-x_{2}-2r_{\emph{b}})\phi^{\emph{L}}_{\psi}(2S)(\emph{x}_{2},b_{2})\\
& +(2\emph{x}_{2}-2+r_{\emph{b}})r_{2}\phi^{t}_{\psi}(2S)(\emph{x}_{2},b_{2})],
\end{split}
\end{equation}
\begin{equation}
\begin{split}
{\cal A}_{b}=&-8\pi\emph{C}_{\emph{F}}f_{\emph{k}}M^{4}_{B_{c}}(1-r^2_{2})\\
& \times \int^{1}_{0}\text{d}\emph{x}_{1}\text{d}\emph{x}_{2}\int^\infty_{0}\emph{b}_{1}\emph{b}_{2}\text{d}\emph{b}_{1}\text{d}\emph{b}_{2}\phi_{B_{c}}(\emph{x}_{1})\\
&\times \emph{E}_{b}(t_{b})\emph{C}_{b}(t_{b})H_{b}(\alpha,\beta_{b},b_{1},b_{2})S_{t}(x_{1})\\
& \times [r^2_{2}(x_{1}-1)-x_{1}]\phi^{\emph{L}}_{\psi}(2S)(\emph{x}_{2},b_{2}),
\end{split}
\end{equation}
\begin{equation}
\begin{split}
{\cal A}_{c}=&-\frac{32\pi\emph{C}_{\emph{F}}}{\sqrt{2N_{c}}}M^{4}_{B_{c}}(1-r^2_{2})\\
& \times \int^{1}_{0}\text{d}\emph{x}_{1}\text{d}\emph{x}_{2}\text{d}\emph{x}_{3}\int^\infty_{0}\emph{b}_{2}\emph{b}_{3}\text{d}\emph{b}_{2}\text{d}\emph{b}_{3}\phi_{B_{c}}(x_{1})\\
& \times \phi^a_{\emph{k}}(x_{3})\emph{E}_{c}(t_{c})\emph{C}_{c}(t_{c})H_{c}(\alpha,\beta_{c},b_{2},b_{3})\\
& \times [(1-r^2_{2})(1-x_{1}-x_{3})\phi^{\emph{L}}_{\psi}(2S)(\emph{x}_{2},b_{2})\\
& +{r}_{2}(x_{2}-x_{1})\phi^{t}_{\psi}(2S)(\emph{x}_{2},b_{2})],
\end{split}
\end{equation}
\begin{equation}
\begin{split}
{\cal A}_{d}=&-\frac{32\pi\emph{C}_{\emph{F}}}{\sqrt{2N_{c}}}M^{4}_{B_{c}}(1-r^2_{2})\\
& \times \int^{1}_{0}\text{d}\emph{x}_{1}\text{d}\emph{x}_{2}\text{d}\emph{x}_{3}\int^\infty_{0}\emph{b}_{2}\emph{b}_{3}\text{d}\emph{b}_{2}\text{d}\emph{b}_{3}\phi_{B_{c}}(x_{1})\\
& \times \phi^a_{\emph{k}}(x_{3})\emph{E}_{d}(t_{d})\emph{C}_{d}(t_{d})H_{d}(\alpha,\beta_{d},b_{2},b_{3})\\
& \times \{ [2x_{1}-x_{2}-x_{3}+(x_{3}-x_{2})r^2_{2}]\phi^{\emph{L}}_{\psi}(2S)(\emph{x}_{2},b_{2})\\
& +{r}_{2}(x_{2}-x_{1})\phi^{t}_{\psi}(2S)(\emph{x}_{2},b_{2})\},
\end{split}
\end{equation}
with $r_{2}=\frac{m_{\psi(2S)}}{M_{B_{c}}}$, and the distribution amplitudes $\phi^{\emph{L},t}(2S)$ can be found in Ref.\cite{r24}.

The amplitude for $B_{c}$$\rightarrow$ $\eta_{c}(2S)$$K$ decay.
\beq
\begin{split}
{\cal A}_{a}=&-i8\pi\emph{C}_{\emph{F}}f_{\emph{k}}M^{4}_{B_{c}}(1-r^2_{2})\\
& \times \int^{1}_{0}\text{d}\emph{x}_{1}\text{d}\emph{x}_{2}\int^\infty_{0}\emph{b}_{1}\emph{b}_{2}\text{d}\emph{b}_{1}\text{d}\emph{b}_{2}\phi_{B_{c}}(\emph{x}_{1})\\
& \times \emph{E}_{a}(t_{a})\emph{C}_{a}(t_{a})H_{a}(\alpha,\beta_{a},b_{1},b_{2})S_{t}(x_{2})\\
& \times \{ [(1-x_{2}-2r_{\emph{b}})\phi^{\upsilon}_{\eta_c}(2S)(\emph{x}_{2},b_{2})\\
& +(2\emph{x}_{2}-2+r_{\emph{b}})r_{2}\phi^{s}_{\eta_c}(2S)(\emph{x}_{2},b_{2})] \},
\end{split}
\eeq

\begin{equation}
\begin{split}
{\cal A}_{b}=&-i8\pi\emph{C}_{\emph{F}}f_{\emph{k}}M^{4}_{B_{c}}(1-r^2_{2})\\
& \times \int^{1}_{0}\text{d}\emph{x}_{1}\text{d}\emph{x}_{2}\int^\infty_{0}\emph{b}_{1}\emph{b}_{2}\text{d}\emph{b}_{1}\text{d}\emph{b}_{2}\phi_{B_{c}}(\emph{x}_{1})\\
&\times \emph{E}_{b}(t_{b})\emph{C}_{b}(t_{b})H_{b}(\alpha,\beta_{b},b_{1},b_{2})S_{t}(x_{1})\\
& \times \{ [r^2_{2}(1-x_{1})+r_{c}]\phi^{\upsilon}_{\eta_c}(2S)(\emph{x}_{2},b_{2})\\
& -2r_{2}[(1-x_{1})+r_{c}]\phi^{\emph{s}}_{\eta_c}(2S)(\emph{x}_{2},b_{2})\},
\end{split}
\end{equation}
\begin{equation}
\begin{split}
{\cal A}_{c}=&-\frac{i32\pi\emph{C}_{\emph{F}}}{\sqrt{2N_{c}}}M^{4}_{B_{c}}(1-r^2_{2})\\
&\times\int^{1}_{0}\text{d}\emph{x}_{1}\text{d}\emph{x}_{2}\text{d}\emph{x}_{3}\int^\infty_{0}\emph{b}_{2}\emph{b}_{3}\text{d}\emph{b}_{2}\text{d}\emph{b}_{3}\phi_{B_{c}}(x_{1})\\
& \times \phi^a_{\emph{k}}(x_{3})\emph{E}_{c}(t_{c})\emph{C}_{c}(t_{c})H_{c}(\alpha,\beta_{c},b_{2},b_{3})\\
& \times \{ [1-x_{1}-x_{3}+r^2_{2}(2x_{2}+x_{3}-x_{1}-1)]\phi^{\upsilon}_{\eta_c}(2S)(\emph{x}_{2},b_{2})\\
& -{r}_{2}(x_{2}-x_{1})\phi^{s}_{\eta_c}(2S)(\emph{x}_{2},b_{2}) \},
\end{split}
\end{equation}
\begin{equation}
\begin{split}
{\cal A}_{d}=&\frac{i32\pi\emph{C}_{\emph{F}}}{\sqrt{2N_{c}}}M^{4}_{B_{c}}(1-r^2_{2})\\
&\times\int^{1}_{0}\text{d}\emph{x}_{1}\text{d}\emph{x}_{2}\text{d}\emph{x}_{3}\int^\infty_{0}\emph{b}_{2}\emph{b}_{3}\text{d}\emph{b}_{2}\text{d}\emph{b}_{3}\phi_{B_{c}}(x_{1})\\
& \times \phi^a_{\emph{k}}(x_{3})\emph{E}_{d}(t_{d})\emph{C}_{d}(t_{d})H_{d}(\alpha,\beta_{d},b_{2},b_{3})\\
& \times \{ [x_{2}+x_{3}-2x_{1}+r^2_{2}(x_{2}-x_{3})]\phi^{\upsilon}_{\eta_c}(2S)(\emph{x}_{2},b_{2})\\
& -{r}_{2}(x_{2}-x_{1})\phi^{s}_{\eta_c}(2S)(\emph{x}_{2},b_{2})\},
\end{split}
\end{equation}
with $r_{2}=\frac{m_{\eta_c(2S)}}{M_{B_{c}}}$, and the distribution amplitudes $\phi^{\upsilon,s}_{\eta_c}(2S)$ have been given in Ref.\cite{r24}.

As for the decay amplitude $B_{c}$$\rightarrow$$\psi(1^3D_1)$$K$, it is similar to the decay amplitude for $B_{c}$$\rightarrow$$\psi(2S)$$K$, but with the replacement $\phi^{\emph{L},t}_{\psi(2s)}$$\rightarrow$$\phi^{\emph{L},t}_{\psi(1D)}$.

The branching ratios for the decay $B_{c}\rightarrow \emph{X}_{c\bar c}K$ in the $B_{c}$ meson rest frame can be written as
\begin{equation}
BR(B_{c}\rightarrow \emph{X}_{c\bar c}K)=\frac{\tau_{B_{c}}}{8\pi}\frac{p}{M^2_{B_{c}}}\mid {\cal A}(B_{c}\rightarrow \emph{X}_{c\bar c}K) \mid^2,
\end{equation}
where the $X_{c\bar c}$ represents the meson $\psi(2S)$, $\eta_{c}(2S)$ and the charmonium state $\psi(1^3D_{1})$ respectively and the common momentum $p=(M^2_{B_{c}}-m^2_{X_{c\bar c}})/{2M_{B_{c}}}$. As for the decay mode $B_{c}\rightarrow \psi(3770)K$, we give the expression based on the idea of S-D mixing scheme:
\begin{equation}
BR(B_{c}\rightarrow \psi(3770)\emph{K})=\frac{\tau_{B_{c}}}{8\pi}\frac{p}{M^2_{B_{c}}}\mid {\cal A}(B_{c}\rightarrow \psi(3770)\emph{K} \mid^2,
\end{equation}
with
\begin{equation}
{\cal A}(B_{c}\rightarrow \psi(3770)\emph{K})=
\cos\theta {\cal A}(B_{c}\rightarrow \psi(1^3D_{1})\emph{K}) - \sin\theta {\cal A}(B_{c}\rightarrow \psi(2S)\emph{K})
\end{equation}

%%%--=================================================================
%%%=====           Numerical evaluation and discussions   ============
%%5===================================================================

\section{Numerical evaluation and discussions}
In our numerical calculation, we adopt the following input parameters\cite{r11,r24,r241,Olive,r40}:
\begin{equation}
\begin{split}
& m_{\psi_{2S}}=3.686{\rm GeV}\;, \quad m_{\eta_{c}(2S)}=3.639{\rm GeV}\;,  \quad M_{B_{c}}=6.277{\rm GeV}\;, \\
& f_{\psi_{2S}}=296^{+3}_{-2}{\rm MeV}\;, \quad f_{B_{c}}=489 \pm 4{\rm MeV}\;, \quad f_{\eta_{c}(2S)}=243^{+79}_{-111}{\rm MeV}\;, \\
& m_{c}=1.28{\rm GeV}\;,  \quad m_{b}=4.18{\rm GeV}\;, \quad m_{\psi_{1D}}=3.77{\rm GeV}\;, \quad a_{1}=0.32\;, \quad a_{2}=-0.34\;\\
& \tau_{B_{c}}=(0.453\pm0.041) {\rm ps}\;, \quad \mid V_{us}\mid=0.2252\pm0.0009\;,  \quad \mid V_{cb}\mid=0.0409\pm0.0011\;.
\end{split}
\end{equation}

If not specified, we will adopt their central values as the default input.

\begin{table}[htbp]
\centering
\caption{The form factors $A_{0,1,2}$ and $V$ of the transition $B_{c}\rightarrow\psi(1^{3}D_{1})$ are calculated in the pQCD approach.}
\label{1D}
\begin{tabular*}{\columnwidth}{@{\extracolsep{\fill}}lllll@{}}
\hline
\hline
            & $A_{0}$              &$A_{1}$   & $A_{2}$  & $V$                \\
\hline
\\
 $\omega_{1D}=0.35$    &4.86$\times10^{-2}$   &10.6$\times10^{-2}$     &0.38     &0.42$\times10^{-2}$  \\
          \\
 $\omega_{1D}=0.40$    &4.97$\times10^{-2}$   &10.0$\times10^{-2}$     &0.33     &2.14$\times10^{-2}$  \\
          \\
 $\omega_{1D}=0.45$    &5.01$\times10^{-2}$   &9.18$\times10^{-2}$     &0.30     &2.50$\times10^{-2}$  \\
          \\
 $\omega_{1D}=0.50$    &4.91$\times10^{-2}$   &8.59$\times10^{-2}$     &0.27     &2.73$\times10^{-2}$   \\
          \\
 $\omega_{1D}=0.55$    &4.46$\times10^{-2}$   &7.98$\times10^{-2}$     &0.25     &2.66$\times10^{-2}$   \\
          \\
 \cite{r18}&2.86$\times10^{-2}$  &4.9$\times10^{-2}$  &0.11 & 0.11      \\
\hline
\end{tabular*}
\end{table}

\begin{table}[htbp]
\centering
\caption{Branching fractions of the decays $B_{c}\rightarrow\emph{X}_{c\bar c}\emph{K}$ in the pQCD approach.}
\label{brch}
\begin{tabular*}{\columnwidth}{@{\extracolsep{\fill}}cccccccc@{}}
\hline
\hline
decay modes                                  &$\omega$    &this work        &\cite{r4} &\cite{r6} &\cite{r8}  &\cite{r9} &\cite{zhu}         \\
\hline
\\
$B_{c}\rightarrow$$\psi(2S)$$ \emph{K}$&0.20&3.33$\times10^{-5}$&(2.30$\pm$0.32)$\times10^{-5}$&1.0$\times10^{-5}$&2.0$\times10^{-5}$&9.3$\times10^{-6}$&5.7$\times10^{-5}$  \\
\\
$B_{c}\rightarrow$$\eta_{c}(2S) $$\emph{K}$   &0.20  & 1.74$\times10^{-5}$ &&1.0$\times10^{-5}$&2.2$\times10^{-5}$&5.0$\times10^{-6}$&     \\
\\
$B_{c}\rightarrow$$\psi(1^3D_{1})$$\emph{K}$  &0.35  & 9.55$\times10^{-6}$ &&&&&      \\
\\
                                              &0.40  & 8.06$\times10^{-6}$ &&&&&      \\
\\
                                              &0.45  & 6.94$\times10^{-6}$ &&&&&      \\
\\
                                              &0.50  & 5.88$\times10^{-6}$ &&&&&      \\
\\
                                              &0.55  & 5.08$\times10^{-6}$ &&&&&      \\
\hline
\end{tabular*}
\end{table}

\begin{table}[htbp]
\centering
\caption{The theoretical uncertainties derived from the decay constant $\emph{f}_{B_{c}}$ and the hard scale $\emph{t}\pm0.15t$, we compute the form factors $A_{0,1,2}$ and $V$ at $q^{2}=0$ for the transition $B_{c}\rightarrow\psi(1^{3}D_{1})$ in the pQCD approach based on the preferred shape paremeter $\omega_{1D}=0.50$GeV.}
\label{1D2}
\begin{tabular*}{\columnwidth}{@{\extracolsep{\fill}}lllll@{}}
\hline
\hline
$\omega_{1D}$    & $A_{0}$                               & $A_{1}$                   \\
          \\
 0.5GeV    &$4.91^{+0.04+0.46}_{-0.04-0.02}$$\times10^{-2}$   &$8.59^{+0.07+0.52}_{-0.07-0.01}$$\times10^{-2}$           \\
           \\
            & 2.86$\times10^{-2}${\cite{r18}}       & 4.9$\times10^{-2}$\cite{r18}        \\
\hline
           & $A_{2}$                                & $V$                                  \\
          \\
 0.5GeV   &$0.27^{+0.00+0.01}_{-0.00-0.00}$       &$2.73^{+0.02+0.58}_{-0.02-0.10}$$\times10^{-2}$  \\
          \\
           & 0.11\cite{r18}                         &0.11\cite{r18}                          \\
\hline
\end{tabular*}
\end{table}

\begin{table}[htbp]
\centering
\caption{The branching fractions of $B_{c}\rightarrow(\psi(2S), \eta_{c}(2S))\emph{K}$ decays with the different theoretical uncertainties arised from the shape parameters, the decay constants $\emph{f}_{B_{c}}$, $\emph{f}_{\psi(2s)}$ or $\emph{f}_{\eta_{c}(2s)}$ and the hard scale $t\pm0.15t$. For $\psi(1^{3}D_{1})$ charmonium state, we adopt the preferred parameter $\omega_{1D}=0.5$Gev. Because there is no uncertainty about the decay constant $\emph{f}_{\psi_{1D}}$, we only give the branching fraction of the decay $B_{c}\rightarrow \psi(1^{3}D_{1})\emph{K}$ with the theoretical uncertainty induced by the decay constant $\emph{f}_{B_{c}}$ and the hard scale $t\pm0.15t$.}
\label{hardt}
\begin{tabular*}{\columnwidth}{@{\extracolsep{\fill}}lllll@{}}
\hline
\hline
  decay modes                                                              \\
\hline
  \\
$B_{c}\rightarrow$$\psi(2S)$$ \emph{K}$  &$3.33^{+0.211+0.054+0.067+0.443}_{-0.232-0.054-0.044-0.127}$$\times10^{-5}$ \\
  \\
$B_{c}\rightarrow$$\eta_{c}(2S) $$\emph{K}$ &$1.74^{+0.144+0.028+1.315+0.173}_{-0.114-0.028-1.227-0.054}$$\times10^{-5}$ \\
  \\
$B_{c}\rightarrow\psi(1^3D_{1})\emph{K}$ &$5.88^{+0.096+0.238}_{-0.095-0.054}$$\times10^{-6}$  \\
  \\
\hline
\end{tabular*}
\end{table}

\begin{table}[htbp]
\centering
\caption{Branching fractions of the decay $B_{c}\rightarrow\psi(3770)\emph{K}$ in the pQCD approach based on two sets of S-D mixing angle.}
\label{3770}
\begin{tabular*}{\columnwidth}{@{\extracolsep{\fill}}lllll@{}}
\hline
\hline
               &$\omega_{1D}$               &$\theta=-12^{\circ}$       &$\theta=27^{\circ}$   \\
\hline
  \\
               &0.35 GeV                    & 1.770$\times10^{-5}$      & 3.012$\times10^{-8}$           \\
  \\
               &0.40 GeV                    & 1.570$\times10^{-5}$      & 1.540$\times10^{-8}$           \\
  \\
               &0.45 GeV                    & 1.414$\times10^{-5}$      & 8.495$\times10^{-8}$           \\
  \\
               &0.50 GeV                    & 1.264$\times10^{-5}$      & 2.113$\times10^{-7}$           \\
    \\
               &0.55 GeV                    & 1.148$\times10^{-5}$      & 3.705$\times10^{-7}$           \\
\hline
\end{tabular*}
\end{table}

Our numerical results for the form factors $A^{B_{c}\rightarrow\psi(1^{3}D_{1})}_{0,1,2}$ and $V^{B_{c}\rightarrow\psi(1^{3}D_{1})}$ are listed in
Table~\ref{1D}, and in Table~\ref{brch} we show the result of the branching fractions for $B_{c}\rightarrow\emph{X}_{c\bar c}\emph{K}$ decays. The branching fractions for $B_{c}\rightarrow\psi(3770)\emph{K}$ decay is displayed in Table~\ref{3770}, which contain two sets of S-D mixing angle. The relative theoretical uncertainties are listed in both Table~\ref{1D2} and Table~\ref{hardt}.

From Table~\ref{1D}, we can find that the form factors of $A_{1,2}$ decrease as the frequency of oscillations parameter $\omega_{1D}$ gets bigger, while $A_{0}$ and $V$ increase first and then decrease. In addition, the form factors are not sensitive to the shape parameter $\omega_{1D}$ except the value of the form factor V at $\omega_{1D}=0.35$ GeV. We also notice that our calculations of form factors are close to the calculations in Ref.\cite{r18} from the light-cone QCD sum rule approach except for the form factor $V$, which is almost four times the result of ours. We expect that it could be compared with more results calculated by other theoretical methods.

From Table~\ref{brch}, we find that our predictions of branching fraction for the decay $B_{c}\rightarrow$$\psi(2S)$$ \emph{K}$ and $B_{c}\rightarrow$$\eta_{c}(2S)$$ \emph{K}$ are close to the predictions in Refs.\cite{r4,r6,r8,zhu}, this indicate the harmonic-oscillator wave functions for radially excited 2S charmonium states is reasonable and applicable. It is clear that the branching fraction of the decay mode $B_{c}\rightarrow\psi(1^3D_{1})\emph{K}$ decrease with the increasing parameter $\omega_{1D}$, and its change trend become slowly as $\omega_{1D}\geq0.5$ GeV, so we take the branching fraction result at $\omega_{1D}=0.5$ GeV as the preferred value. In view of the success about the B meson exclusive decay $B\rightarrow$$\psi(3770)$$ \emph{K}$\cite{r41}, we calculate the branching fraction of $B_{c}\rightarrow$$\psi(3770)$$ \emph{K}$ based on the S-D mixing scheme, whose computations are listed in Table~\ref{3770}, here the selection basis of the S-D mixing angle has been introduced in Sect.I. By comparing Table~\ref{brch} and Table~\ref{3770}, we find that, for the decay $B_{c}\rightarrow$$\psi(3770)$ $\emph{K}$ near our preferred shape parameter $\omega_{1D}=0.5$ GeV, we get a branching fraction in order of magnitude $10^{-6}$ if we treat the $\psi(3770)$ as a pure 1D charmonium state. But, the branching fraction can be raised from 5.88$\times10^{-6}$ to 1.264$\times10^{-5}$ obviously when we adopt the mixing angle $\theta=-12^{\circ}$, which is consistent with the arguments about mixing angle $\theta$ in Refs.\cite{r17,r18,r41,r121,r122}. We attribute this remarkable improvement to a very small decay constant of charmonium state $\psi(1^{3}D_{1})$ (0.0478 GeV) compared with the decay constant of the meson $\psi(2S)$ (0.296 GeV). This decay mode has not been measured yet, but around $\cal{O}$$(10^{9})$ $B_{c}$ mesons can be anticipated with 1 $fb^{-1}$ of data at the LHC \cite{YNGao}, which
make it could be soon tested at the LHC-b experiment, that will help us to understand the structure of $\psi(3770)$ and the constituent quark model.

Moreover, we give the relative theoretical uncertainties in Table~\ref{1D2} and Table~\ref{hardt}. In Table~\ref{1D2}, we analyzed the uncertainties of $B_{c}\rightarrow\psi(1^{3}D_{1})$ transition form factors based on the preferred value $\omega_{1D}=0.5$ GeV. The two theoretical uncertainties come from the decay constant $\emph{f}_{B_{c}}$ and the hard scale $\emph{t}\pm0.15t$ in Eq. (\ref{eq:exp5}), which characterizes the size of next to leading order contribution. From Table~\ref{1D2} and Table~\ref{1D}, it is easy to see that the main error come from the nonperturbative shape parameter, which need more theoretical and experimental efforts to
understand. In Table~\ref{hardt}, we display the branching fractions of $B_{c}$ $\rightarrow$($\psi(2S)$, $\eta_{c}(2S)$, $\psi(1^3D_{1})$)$\emph{K}$ decays with different theoretical uncertainties. For the decay mode $B_{c}\rightarrow$$\psi(2S)$$\emph{K}$, the main theoretical uncertainties come from the shape parameter error $\omega_{c}=0.2\pm0.01$ for $\psi(2S)$ meson and hard scale $t\pm0.15t$, which produce an uncertainty in the range of -6.9$\%$ to 6.3$\%$ and -3.8$\%$ to 13$\%$ respectively. For the decay mode $B_{c}\rightarrow$$\eta_{c}(2S)$$\emph{K}$, the shape parameter error produce an uncertainty in the range of from $-6.5\%$ to 8.3$\%$ and the hard scale $t$ leads to an uncertainty from $-3.1\%$ to 9.9$\%$. For the decay $B_{c}\rightarrow\psi(1^3D_{1})\emph{K}$, the larger uncertainty coming from the hard scale $\emph{t}$ which can bring an uncertainty of -0.9$\%$ to 4.0$\%$ to the branching ratio. These small uncertainties show that the harmonic oscillator wave function is an excellent candidate for describing charmonium states and the uncertainties from the next to leading order contributions are very limited and can be neglected safely for these decay modes. The largest uncertainty of the decay $B_{c}\rightarrow$$\eta_{c}(2S)$$\emph{K}$ appears in the decay constant $f_{\eta_{c}(2s)}$, this point is easy to understand for the great uncertainty $f_{\eta_{c}(2S)}=243^{+79}_{-111}$MeV, whose origin have been studied in Ref.\cite{r24}, and whose value is expected to be improved by future precise experimental measurements at LHC-b or Super-B factories. The other uncertain factors such as CKM matrix elements and $B_{c}$ meson life are too small and can be neglected safely.

\section{Summary} \label{sec:summary}

In this paper, we calculated the form factors of $B_{c}\rightarrow \psi(1^3D_{1})$ and gave the predictions for the branching fractions of two-body decays $B_{c}\rightarrow$$\psi(2S)$ $\emph{K}$, $\eta_{c}(2S)$$\emph{K}$, $\psi(3770)$$\emph{K}$ in the perturbative QCD approach. The new orbitally excited charmonium distribution amplitudes of $\psi(1^{3}D_{1})$ based
on the Schr\"{o}dinger wave function of the $n=1$, $l=2$ state for the harmonic-oscillator potential are employed. We also discussed the theoretical uncertainties in this paper.
 In view of the mixing mechanism of S-D wave, we gave a theoretical calculation of the branching ratio for the decay $B_{c}\rightarrow$$\psi(3770)$ $\emph{K}$ firstly in the literature. Our calculations show that the branching ratio of the decay $B_{c}$$\rightarrow$$\psi(3770)$ K can reach the order of $10^{-5}$, which can be tested by the running LHC-b experiments.

%%%--=================================================================
%%%=====            Acknowledgements        ==========================
%%5===================================================================

\begin{acknowledgments}

The authors would to thank
 Ming-Zhen Zhou, Jun-Feng Sun and Xin Liu for some valuable discussions.
This work is supported by the National Natural Science Foundation of
 China under Grant Nos.11047028 and 11645002, and by the Fundamental
 Research Funds of the Central Universities, Grant Number
 XDJK2012C040.

\end{acknowledgments}

%%%--=================================================================
%%%=====            Appendix       ==========================
%%5===================================================================
\section{Appendix : Formulas For The Calculation Used In The Text} \label{sec:appendix}

The function $E_{i}(t_{i})$ in the decay amplitudes are defined by
\begin{equation}
E_{i}(t_{i})=\alpha_{s}(t_{i})e^{-S_{i}},
\end{equation}
in which the strong running coupling constant $\alpha_{s}(t_{i})$\cite{r42} based on the standard one-loop calculations is adopted in our calculations and
\begin{equation}
\begin{split}
&S_{a(b)}=S_{B_{c}}(t_{a(b)})+S_{X_{c\bar c}}(t_{a(b)})=s(x_{1}p^+_{1},b_{1})+s(x_{2}p^+_{2},b_{2})+s(\bar{x}_{2}p^+_{2},b_{2})\\
&+2\int^{t}_{1/b_{1}}\frac{\text{d}\mu}{\mu}\gamma_{q}+2\int^{t}_{1/b_{2}}\frac{\text{d}\mu}{\mu}\gamma_{q}\\
&S_{c(d)}=S_{B_{c}}(t_{a(b)})+S_{X_{c\bar c}}(t_{c(d)})+S_{\emph{K}}(t_{c(d)})=s(x_{1}p^+_{1},b_{2})+s(x_{2}p^+_{2},b_{2})+s(\bar{x}_{2}p^+_{2},b_{2})+s(x_{3}p^-_{3},b_{3})\\
&+s(\bar{x}_{3}p^-_{3},b_{3})+2\int^{t}_{1/b_{2}}\frac{\text{d}\mu}{\mu}\gamma_{q}+2\int^{t}_{1/b_{2}}\frac{\text{d}\mu}{\mu}\gamma_{q}+2\int^{t}_{1/b_{3}}\frac{\text{d}\mu}{\mu}\gamma_{q}\\
\end{split}
\end{equation}
where the functions $s(Q,b)$ are called Sudakov factor resulting from the resummation of double logarithms and can be found in \cite{r43}. $\gamma_{q}=-\alpha_{s}/\pi$ is the anomalous dimension of the quark.
For killing the large logarithmic radiative corrections, the hard scale $t_{i}$ in the amplitudes are choosen as the maximum in $H_{i}$ \begin{equation}
\begin{split}
&t_{a(b)}=max(\sqrt{|\alpha|},\sqrt{|\beta_{a(b)}|},1/b_{1},1/b_{2}),\\
&t_{c(d)}=max(\sqrt{|\alpha|},\sqrt{|\beta_{c(d)}|},1/b_{2},1/b_{3}).\\
\end{split}
\end{equation}
The hard scattering kernels function H arising from the Fourier transform of virtual quark and gluon propagators and are written as follows
\begin{equation}
\begin{split}
&H_{i}(\alpha,\beta_{i},b_{1},b_{2})=H_{1}(\alpha,b_{1})\times H_{2}(\beta_{i},b_{1},b_{2})\\
&H_{1}(\alpha,b_{1})=
\begin{cases}
K_{0}(b_{1}\sqrt{\alpha})& {\alpha > 0}\\
K_{0}(ib_{1}\sqrt{-\alpha})& {\alpha < 0}\\
\end{cases}\\
&H_{2}(\beta_{i},b_{1},b_{2})=\\
&\begin{cases}
\theta(b_{1}-b_{2})K_{0}(b_{1}\sqrt{\beta})I_{0}(b_{2}\sqrt{\beta})+(b_{1}\leftrightarrow b_{2})& {\beta > 0}\\
\theta(b_{1}-b_{2})K_{0}(ib_{1}\sqrt{-\beta})J_{0}(b_{2}\sqrt{-\beta})+(b_{1}\leftrightarrow b_{2})& {\beta < 0}\\
\end{cases}\\
\end{split}
\end{equation}
\begin{equation}
\begin{split}
&\alpha=-M^{2}_{B_{c}}(x_{1}-x_{2})(x_{1}-r^{2}_{2}x_{2}),\\
&\beta_{a}=-M^{2}_{B_{c}}[(1-x_{2})(1-r^{2}_{2}x_{2})-r^{2}_{b}],\\
&\beta_{b}=-M^{2}_{B_{c}}[(1-x_{1})(r^{2}_{2}-x_{1})-r^{2}_{c}],\\
&\beta_{c}=-M^{2}_{B_{c}}(x_{2}-x_{1})[(\bar{x}_{3}-x_{1})+(x_{2}-\bar{x}_{3})r^{2}_{2}],\\
&\beta_{d}=-M^{2}_{B_{c}}(x_{2}-x_{1})[(x_{3}-x_{1})+(x_{2}-x_{3})r^{2}_{2}],\\
\end{split}
\end{equation}
\begin{equation}
C_{a(b)}=C_{1}+C_{2}/N_{c},C_{c(d)}=C_{2},
\end{equation}
where $J_{0}$ is the Bessel function and $K_{0}$, $I_{0}$ are modified Bessel function with $K_{0}(ix)=\frac{\pi}{2}(-N_{0}(x)+iJ_{0}(x))$. The $C_{1,2}$ are the Wilson coefficients.

The jet function $S_{t}(x)$ coming from the threshold resummation\cite{r32} contribute to the factorizable diagrams (a) and (b) in Fig.~\ref{fig:fig2}

\begin{equation}
S_{t}(x)=\frac{2^{1+2c}\Gamma(\frac{3}{2}+c)}{\sqrt{\pi}\Gamma(1+c)}[x(1-x)]^{c},c=0.3
\end{equation}

%%%--=================================================================
%%%=====           Reference        ==========================
%%5===================================================================

\end{document}